\documentstyle[multicol,prb,aps,ifthen,epsfig]{revtex}

\newcommand{\wide}[2]{                                                        %
\end{multicols}                                                               %
\widetext                                                                     %
\noindent                                                                     %
\ifthenelse{\equal{#1}{t}}                                                    %
{}                                                                            %
{                                                                             %
\raisebox{0.1in}[0in][0.02in]{$\rule{3.575in}{0.002in}                        %
\rule{0.002in}{0.08in}$}                                                      %
}                                                                             %
#2                                                                            %
\ifthenelse{\equal{#1}{b}}                                                    %
{}                                                                            %
{                                                                             %
{\raisebox{-0.1in}[0in][0.02in]                                               %
{\hspace{3.575in}$\rule{0.002in}{0.08in}                                      %
\rule[0.08in]{3.575in}{0.002in}$}                                             %
}                                                                             %
}                                                                             %
\begin{multicols}{2}                                                          %
\noindent                                                                     %
}                                                                             %
\def  \no     {\nonumber}

\begin{document}

\title{Effective resistivity of magnetic multilayers}
\author{A. Cr\'epieux* and P. Bruno}
\address{Max-Planck-Institut f\"{u}r Mikrostrukturphysik, Weinberg
2, 06120 Halle, Germany}
\date{\today}
\maketitle

\begin{abstract}
In heterogeneous system, the correspondence between calculated
and measured quantities, such as the conductivity or the
resistivity, is not obvious since the former ones are local
quantities whereas the latter ones are often average values over
the sample. In this report, we show explicitly how the
correspondence can be done in the case of magnetic multilayers.
\end{abstract}

\begin{multicols}{2}
In the linear response regime, the electric current ${\bf J}$ at
position ${\bf r}$ is related to the electric field ${\bf E}$ at
position ${\bf r}'$ through\cite{Mahan}
\begin{eqnarray}\label{linear}
  J_i({\bf r})=\int_{-\infty}^{+\infty} d{\bf r}'\sum_{j}\sigma_{ij}({\bf r},{\bf r}')E_j({\bf r}'),
\end{eqnarray}
where $i$ and $j$ refer to the space directions $\{x,y,z\}$. The
conductivity $\sigma_{ij}$ depends of both ${\bf r}$ and ${\bf
r}'$ (two-points conductivity), it is a local quantity which is
expressed in the Kubo formalism\cite{Kubo} as a current-current
correlation function
\begin{eqnarray}\label{correlation}
  \sigma_{ij}({\bf r},{\bf r}')=\lim_{\omega\rightarrow
  0}\frac{1}{\omega}\int_{0}^{+\infty}dt\;e^{i\omega t}
  \left\langle\;\left[\;{\bf j}({\bf r},t)\;;\;{\bf j}({\bf r}',0)\;\right]\;\right\rangle,
\end{eqnarray}
where ${\bf j}$ is the current density and the bracket
$[\;A\;;\;B\;] = AB - BA$. The notation
$\langle\cdot\cdot\cdot\rangle$ refers to the configurational
average. The question we address in this report is how to link
the calculated quantities given by Eq.~(\ref{correlation}) which
are local (position dependent) with the measured quantities which
are mostly non-local. Indeed, the size of the electric contacts
used to measure the conductivity (by measuring the current and
the voltage) are generally quite large (from $\mu$m to mm). Then
the measured quantity is a kind of average over a part of the
sample. Strictly speaking, it is rather  an effective value than
an average value. The relation between the measured and
calculated quantities varies strongly
with the geometry of the system.

Let us start with a three-dimensional homogeneous system in the stationary regime. In this
case, the electric current and electric field do not dependent of
the position, thus Eq.~(\ref{linear}) reduces to
$J_i=\sum_{j}\bar{\sigma}_{ij}E_j$ where we have introduced the
effective conductivity
\begin{eqnarray}\label{3d}
  \bar{\sigma}_{ij}\equiv\int_{-\infty}^{+\infty} d{\bf r}'\;\sigma_{ij}({\bf r},{\bf
  r}'),
\end{eqnarray}
which is nothing else than the average value over the sample. The
${\bf r}$-dependence in the r.s.h. of Eq.~(\ref{3d}) is
irrelevant since the system is invariant by translation. Then, for
an homogeneous system, the calculated (two-points conductivity
$\sigma_{ij}({\bf r},{\bf
  r}')$) and the measured (effective conductivity $\bar{\sigma}_{ij}$) quantities
are related by a simple relation.

It is not more the case for heterogeneous systems because the
electric current and electric field are not uniform through the
sample. In this report, we treat the case of a multilayer formed
by a superposition of identical cells which have the dimension
$\{a_0,a_0,La_0\}$ where $a_0$ is the interatomic distance and
$L$ the number of layers in the cell (see Fig.~(1a)). We have
still an invariance by translation in each direction but the
periodicity is different in comparison to the three-dimensional
homogeneous system. The indication ${\bf r}$ of the position of
one site must be replaced by the indication ${\bf R}$ of the
position of the cell plus the indication $l$ of the position of
the site in the cell ($l$ varies from $1$ to $L$). For a
multilayer, it is more appropriate to write Eq.~(\ref{linear})
under the form
\begin{eqnarray}\label{linearmultilayer}
  J_i^{l_1}({\bf R})=\int_{-\infty}^{+\infty} d{\bf R}'
  \sum_{j,l_2}\sigma_{ij}^{l_1l_2}({\bf R},{\bf R}')E_j^{l_2}({\bf R}'),
\end{eqnarray}
because we have a translational invariance with respect to the
vectors ${\bf R}$ and ${\bf R}'$, thus the electric current and
electric field do not depend of the cell position. Following the
same procedure than before, we write
\begin{eqnarray}\label{current}
  J_i^{l_1}=\sum_{j,l_2}\sigma_{ij}^{l_1l_2}E_j^{l_2},
\end{eqnarray}
where
\begin{eqnarray}
  \sigma_{ij}^{l_1l_2}\equiv\int_{-\infty}^{+\infty}d{\bf R}'\;\sigma_{ij}^{l_1l_2}({\bf R},{\bf
  R}').
\end{eqnarray}
However, we can not design this quantity as the effective
conductivity since it is a local quantity through the $l_1$ and
$l_2$ indices. Since a multilayer is formed by a succession of
layers made of different material, it is not possible to assume
that the electric current and electric field are layer
independent. To get the effective conductivity, we have to go
further. In the literature, only approximate expressions have
been proposed\cite{Camblong,Levy,Butler}. They are controlled by
the relative value of two lengths: the mean-free-path $\lambda$
and the average thickness d of the uniform layers in the cell.
When $\lambda\gg d$ (homogeneous limit), the electrons can go
through many layers without experience any scattering. They do
not feel the difference between the layers and thus the cell can
be considered as an homogeneous system. The layer dependence of
the conductivity is not relevant and the multilayer can be
treated as a three-dimensional homogeneous system (see
Eq.~(\ref{3d})). When $\lambda\ll d$ (local limit), the electrons
are scattered many times before they go out of one layer: the
current in one layer depends only weakly of the electric field in
the other layers. As a consequence, the contributions of the
two-points conductivity ${\sigma}_{ij}^{l_1\neq l_2}$ are
negligible, and Eq.~(\ref{current}) reduces to
$J_i^l=\sum_{j}{\sigma}_{ij}^{ll}E_j^l$ which is still layer
dependent. Thus, in the local limit, the effective conductivity
can not be directly obtained. A general derivation of the exact
effective conductivity valid in any regimes would be more
appropriate. In this report, we present such a derivation. The
starting point is Eq.~(\ref{current}) where the two-points
conductivity $\sigma_{ij}^{l_1l_2}$ links the local electric
current $J_i^{l_1}$ to the local electric field $E_j^{l_2}$. The
conductivity is a $(3L\times 3L)$ matrix. By inversion of this
matrix, we get the matrix elements $\rho_{ij}^{l_1l_2}$
(two-points resistivity) which link the local electric field
$E_i^{l_1}$ to the local electric current $J_j^{l_2}$
\begin{eqnarray}\label{field}
  E_i^{l_1}=\sum_{j,l_2}\rho_{ij}^{l_1l_2}J_j^{l_2}.
\end{eqnarray}
The important thing now is to consider the experiments, in
particular the geometry of the system and the way how the
contacts are made. We distinguish two different geometries :
the CIP geometry (current in the plane of the layers, see Fig.~(1b))
and the CPP geometry (current perpendicular to the plane of the layers,
see Fig.~(1c)).
\begin{figure}
\centering \epsfig{file=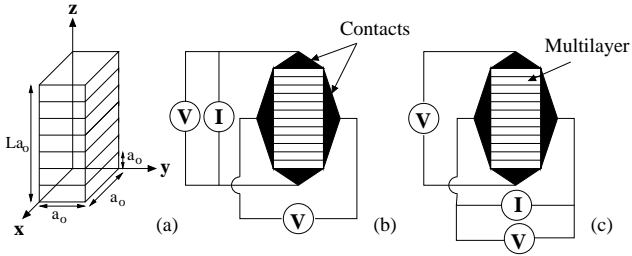}\caption{(a) Schema of the unit
cell for a multilayer; (b)~CIP~geometry; (c) CPP geometry.}
\end{figure}
In CPP geometry, the current is perpendicular to the layers. Due
to current conservation, the perpendicular current is uniform
through the cell (i.e., $J_z^l=J_z$) and the measured value is
$J_z$. In CIP geometry, the current is parallel to the layers.
Since the layers are different, the in-plane currents $J_x^l$ and
$J_y^l$ are layer dependent. As it is shown in Fig~(1c), the
contacts cover several layers. We can assume that they cover at
least a whole cell and that they do not have any influence on it,
then what is measured are the average currents
$\overline{J}_x\equiv\sum_{l}J_x^{l}/L$ and
$\overline{J}_y\equiv\sum_{l}J_y^{l}/L$. For both geometries, the
electrical field can have parallel and perpendicular components.
Since the layers are different, the three components of the
electric field should be layer dependent. However, the Maxwell
equation in the stationary regime ($\nabla\times{\bf E}=0$)
combined with the fact that the electric field component $E_z^l$
is uniform in the xy-plane impose that in-plane components of
${\bf E}$ are layer independent ($E_x^l=E_x$ and $E_y^l=E_y$).
Thus, $E_x$ and $E_y$ correspond to the measured quantities. On
the contrary, $E_z^l$ stays layer dependent and what is measured
is the average electric field
$\overline{E}_z\equiv\sum_{l}E_z^{l}/L$.

By means of some
transformations on Eqs.~(\ref{current}) and (\ref{field}), we
express $J_x^{l}$, $J_y^{l}$ and $E_z^{l}$ with the help of the
uniform quantities $E_x$, $E_y$ and $J_z$. For the moment, all
the indices are kept. Eq.~(\ref{current}) can be written as
\begin{eqnarray}\label{eq1}
  \sum_{l_2}\sigma_{zz}^{l_1l_2}E_z^{l_2}=-\sum_{l_2}\sigma_{zx}^{l_1l_2}E_x^{l_2}
  -\sum_{l_2}\sigma_{zy}^{l_1l_2}E_y^{l_2}+J_z^{l_1}.
\end{eqnarray}
>From Eq.~(\ref{field}), we have
\begin{eqnarray}\label{eq2}
  \sum_{l_2}\rho_{xx}^{l_1l_2}J_x^{l_2}
  +\sum_{l_2}\rho_{xy}^{l_1l_2}J_y^{l_2}=E_x^{l_1}-\sum_{l_2}\rho_{xz}^{l_1l_2}J_z^{l_2},
\end{eqnarray}
and
\begin{eqnarray}\label{eq3}
  \sum_{l_2}\rho_{yx}^{l_1l_2}J_x^{l_2}
  +\sum_{l_2}\rho_{yy}^{l_1l_2}J_y^{l_2}=E_y^{l_1}-\sum_{l_2}\rho_{yz}^{l_1l_2}J_z^{l_2}.
\end{eqnarray}
These three equations can be written under the matrix form
\begin{eqnarray}\label{sys1}
    &&\left(\begin{array}{ccc}
    \tilde{\rho}_{xx} & \tilde{\rho}_{xy} & 0 \\
    \tilde{\rho}_{yx} & \tilde{\rho}_{yy} & 0 \\
    0 &  0 & \tilde{\sigma}_{zz}
    \end{array}\right)
    \left(\begin{array}{lll}
    \tilde{J}_x \\
    \tilde{J}_y \\
    \tilde{E}_z
    \end{array}\right)\no\\
    &&=\left(\begin{array}{ccc}
    \tilde{I} & 0 & -\tilde{\rho}_{xz} \\
    0 & \tilde{I} & -\tilde{\rho}_{yz} \\
    -\tilde{\sigma}_{zx} &  -\tilde{\sigma}_{zy} & \tilde{I}
    \end{array}\right)
    \left(\begin{array}{lll}
    \tilde{E}_x \\
    \tilde{E}_y \\
    \tilde{J}_z
    \end{array}\right),
\end{eqnarray}
where $\tilde{I}$ is the $(L\times L)$ identity matrix and where
we have introduced the following notations: the L-components
vector $\tilde{A}_i$ ($A=J$ or $E$) and the $(L\times L)$ matrix
$\tilde{a}_{ij}$ ($a=\sigma$ or $\rho$) define as
\begin{eqnarray}\label{vect}
    \tilde{A}_i\equiv
    \left(\begin{array}{lll}
    A_i^1 \\
    \vdots \\
    A_i^L
    \end{array}\right),\hspace{1cm}
    \tilde{a}_{ij}\equiv\left(\begin{array}{ccc}
    a_{ij}^{11} & \ldots & a_{ij}^{1L} \\
    \vdots & & \vdots \\
    a_{ij}^{L1} & \ldots & a_{ij}^{LL} \\
    \end{array}\right).
\end{eqnarray}
From Eq.~(\ref{sys1}), we get
\begin{eqnarray}\label{sys2}
    \left(\begin{array}{lll}
    \tilde{J}_x \\
    \tilde{J}_y \\
    \tilde{E}_z
    \end{array}\right)
    &=&\left(\begin{array}{ccc}
    \tilde{\rho}_{xx} & \tilde{\rho}_{xy} & 0 \\
    \tilde{\rho}_{yx} & \tilde{\rho}_{yy} & 0 \\
    0 &  0 & \tilde{\sigma}_{zz}
    \end{array}\right)^{-1}\no\\
    &&\times\left(\begin{array}{ccc}
    \tilde{I} & 0 & -\tilde{\rho}_{xz} \\
    0 & \tilde{I} & -\tilde{\rho}_{yz} \\
    -\tilde{\sigma}_{zx} &  -\tilde{\sigma}_{zy} & \tilde{I}
    \end{array}\right)
    \left(\begin{array}{lll}
    \tilde{E_x} \\
    \tilde{E}_y \\
    \tilde{J}_z
    \end{array}\right).
\end{eqnarray}
We inverse the first matrix in the r.s.h (it can be done
numerically) and perform the multiplication of the two matrices,
thus we get

\begin{eqnarray}\label{sys3}
    \left(\begin{array}{lll}
    \tilde{J}_x \\
    \tilde{J}_y \\
    \tilde{E}_z
    \end{array}\right)
    &=&\left(\begin{array}{ccc}
    \tilde{s}_{xx} & \tilde{s}_{xy} & -(\tilde{s}_{xx}\tilde{\rho}_{xz}+\tilde{s}_{xy}\tilde{\rho}_{yz}) \\
    \tilde{s}_{yx} & \tilde{s}_{yy} & -(\tilde{s}_{yx}\tilde{\rho}_{xz}+\tilde{s}_{yy}\tilde{\rho}_{yz}) \\
    -\tilde{p}_{zz}\tilde{\sigma}_{zx} & -\tilde{p}_{zz}\tilde{\sigma}_{zy} & \tilde{p}_{zz} \\
    \end{array}\right)
\no\\
    &&\times\left(\begin{array}{lll}
    \tilde{E}_x \\
    \tilde{E}_y \\
    \tilde{J}_z
    \end{array}\right),
\end{eqnarray}
where $\tilde{s}_{ij}$ and $\tilde{p}_{zz}$ are $(L\times L)$
matrices defined through
\begin{eqnarray}\label{inv}
    \left(\begin{array}{ccc}
    \tilde{s}_{xx} & \tilde{s}_{xy} & 0 \\
    \tilde{s}_{yx} & \tilde{s}_{yy} & 0 \\
    0 &  0 & \tilde{p}_{zz}
    \end{array}\right)
    \equiv\left(\begin{array}{ccc}
    \tilde{\rho}_{xx} & \tilde{\rho}_{xy} & 0 \\
    \tilde{\rho}_{yx} & \tilde{\rho}_{yy} & 0 \\
    0 &  0 & \tilde{\sigma}_{zz}
    \end{array}\right)^{-1}.
\end{eqnarray}
By using the fact that $E_x$, $E_y$ and $J_z$ are
layer-independent, we can sum over the columns and reduce the
initial size $(3L \times 3L)$ of the matrix which appears in
Eq.~(\ref{sys3}) to the size $(3L \times 3)$
\wide{m}{
\begin{eqnarray}\label{sys4}
    \left(\begin{array}{lllllllll}
    J_x^1 \\
    \vdots \\
    J_x^L \\
    J_y^1 \\
    \vdots \\
    J_y^L \\
    E_z^1 \\
    \vdots \\
    E_z^L
    \end{array}\right)=
    \left(\begin{array}{ccccccccc}
    \sum_{l_2}s_{xx}^{1l_2}&\sum_{l_2}s_{xy}^{1l_2}&-\sum_{l_2,l_3}\left(s_{xx}^{1l_3}\rho_{xz}^{l_3l_2}+s_{xy}^{1l_3}\rho_{yz}^{l_3l_2}\right)\\
    \vdots&\vdots&\vdots\\
    \sum_{l_2}s_{xx}^{Ll_2}&\sum_{l_2}s_{xy}^{Ll_2}&-\sum_{l_2,l_3}\left(s_{xx}^{Ll_3}\rho_{xz}^{l_3l_2}+s_{xy}^{Ll_3}\rho_{xy}^{l_3l_2}\right)\\
    \sum_{l_2}s_{yx}^{1l_2}&\sum_{l_2}s_{yy}^{1l_2}&-\sum_{l_2,l_3}\left(s_{yx}^{1l_3}\rho_{xz}^{l_3l_2}+s_{yy}^{1l_3}\rho_{yz}^{l_3l_2}\right)\\
    \vdots&\vdots&\vdots\\
    \sum_{l_2}s_{yx}^{Ll_2}&\sum_{l_2}s_{yy}^{Ll_2}&-\sum_{l_2,l_3}\left(s_{yx}^{Ll_3}\rho_{xz}^{l_3l_2}+s_{yy}^{Ll_3}\rho_{yz}^{l_3l_2}\right)\\
    -\sum_{l_2,l_3}p_{zz}^{1l_3}\sigma_{zx}^{l_3l_2}&-\sum_{l_2,l_3}p_{zz}^{1l_3}\sigma_{zy}^{l_3l_2}&\sum_{l_2}p_{zz}^{1l_2}\\
    \vdots&\vdots&\vdots\\
    -\sum_{l_2,l_3}p_{zz}^{Ll_3}\sigma_{zx}^{l_3l_2}&-\sum_{l_2,l_3}p_{zz}^{Ll_3}\sigma_{zy}^{l_3l_2}&\sum_{l_2}p_{zz}^{Ll_2}
    \end{array}\right)
    \left(\begin{array}{lll}
    E_x \\
    E_y \\
    J_z
    \end{array}\right).
\end{eqnarray}
As we are interested by the average currents
$\overline{J}_x\equiv\sum_{l}J_x^{l}/L$,
$\overline{J}_y\equiv\sum_{l}J_y^{l}/L$ and the average electric
field $\overline{E}_z\equiv\sum_{l}E_z^{l}/L$ (it is what we can
get experimentally in such a system), we reduce the $(3L \times
3)$ matrix which appears in Eq.~(\ref{sys4}) to a $(3 \times 3)$
matrix
\begin{eqnarray}\label{sys6}
    \left(\begin{array}{lll}
    \overline{J}_x \\
    \overline{J}_y \\
    \overline{E}_z
    \end{array}\right)=
    \left(\begin{array}{ccc}
    \|{s}_{xx}\|&\|{s}_{xy}\|&-\|{s}_{xx}{\rho}_{xz}+{s}_{xy}{\rho}_{yz}\|\\
    \|{s}_{yx}\|&\|{s}_{yy}\|&-\|{s}_{yx}{\rho}_{xz}+{s}_{yy}{\rho}_{yz}\|\\
    -\|{p}_{zz}{\sigma}_{zx}\|&-\|{p}_{zz}{\sigma}_{zy}\|&\|{p}_{zz}\|\\
    \end{array}\right)
    \left(\begin{array}{lll}
    E_x \\
    E_y \\
    J_z
    \end{array}\right),
\end{eqnarray}
where we have introduced the definition $\|{a}_{ij}\|\equiv
\sum_{l_1,l_2}a_{ij}^{l_1l_2}/L$ in order to simplify the
notations. This system of equations can be written under the form
\begin{eqnarray}\label{sys7}
    \left(\begin{array}{ccc}
    1&0&\|{s}_{xx}{\rho}_{xz}+{s}_{xy}{\rho}_{yz}\|\\
    0&1&\|{s}_{yx}{\rho}_{xz}+{s}_{yy}{\rho}_{yz}\|\\
    0&0&-\|{p}_{zz}\|\\
    \end{array}\right)
    \left(\begin{array}{lll}
    \overline{J}_x \\
    \overline{J}_y \\
    J_z
    \end{array}\right)=
    \left(\begin{array}{ccc}
    \|{s}_{xx}\|&\|{s}_{xy}\|&0\\
    \|{s}_{yx}\|&\|{s}_{yy}\|&0\\
    -\|{p}_{zz}{\sigma}_{zx}\|&-\|{p}_{zz}{\sigma}_{zy}\|&-1\\
    \end{array}\right)
    \left(\begin{array}{lll}
    E_x \\
    E_y \\
    \overline{E}_z
    \end{array}\right).
\end{eqnarray}
}
From Eq.~(\ref{sys7}), the effective conductivity tensor
$\tilde{\sigma}$ is
\begin{eqnarray}\label{sys8}
    \overline{\sigma}&=&
    \left(\begin{array}{ccc}
    1&0&\|{s}_{xx}{\rho}_{xz}+{s}_{xy}{\rho}_{yz}\|\\
    0&1&\|{s}_{yx}{\rho}_{xz}+{s}_{yy}{\rho}_{yz}\|\\
    0&0&-\|{p}_{zz}\|\\
    \end{array}\right)^{-1}\no\\
    &&\times\left(\begin{array}{ccc}
    \|{s}_{xx}\|&\|{s}_{xy}\|&0\\
    \|{s}_{yx}\|&\|{s}_{yy}\|&0\\
    -\|{p}_{zz}{\sigma}_{zx}\|&-\|{p}_{zz}{\sigma}_{zy}\|&-1\\
    \end{array}\right).
\end{eqnarray}
The inversion of the first matrix and the product of the two
matrices lead to the following expressions of the matrix elements
for the effective conductivity
\begin{eqnarray}\label{effcond}
    \overline{\sigma}_{xx}&=&\|{s}_{xx}\|-\frac{\|{s}_{xx}{\rho}_{xz}+{s}_{xy}{\rho}_{yz}\|\|{p}_{zz}{\sigma}_{zx}\|}{\|{p}_{zz}\|},\no\\
    \overline{\sigma}_{xy}&=&\|{s}_{xy}\|-\frac{\|{s}_{xx}{\rho}_{xz}+{s}_{xy}{\rho}_{yz}\|\|{p}_{zz}{\sigma}_{zy}\|}{\|{p}_{zz}\|},\no\\
    \overline{\sigma}_{xz}&=&-\frac{\|{s}_{xx}{\rho}_{xz}+{s}_{xy}{\rho}_{yz}\|}{\|{p}_{zz}\|},\no\\
    \overline{\sigma}_{yx}&=&\|{s}_{yx}\|-\frac{\|{s}_{yx}{\rho}_{xz}+{s}_{yy}{\rho}_{yz}\|\|{p}_{zz}{\sigma}_{zx}\|}{\|{p}_{zz}\|},\no\\
    \overline{\sigma}_{yy}&=&\|{s}_{yy}\|-\frac{\|{s}_{yx}{\rho}_{xz}+{s}_{yy}{\rho}_{yz}\|\|{p}_{zz}{\sigma}_{zy}\|}{\|{p}_{zz}\|},\no\\
    \overline{\sigma}_{yz}&=&-\frac{\|{s}_{yx}{\rho}_{xz}+{s}_{yy}{\rho}_{yz}\|}{\|{p}_{zz}\|},\no\\
    \overline{\sigma}_{zx}&=&\frac{\|{p}_{zz}{\sigma}_{zx}\|}{\|{p}_{zz}\|},\no\\
    \overline{\sigma}_{zy}&=&\frac{\|{p}_{zz}{\sigma}_{zy}\|}{\|{p}_{zz}\|},\no\\
    \overline{\sigma}_{zz}&=&\frac{1}{\|{p}_{zz}\|}.
\end{eqnarray}
From Eq.~(\ref{sys7}), the effective resistivity tensor
$\overline{\rho}$ is
\begin{eqnarray}\label{sys9}
    \overline{\rho}
    =\left(\begin{array}{ccc}
    \|{s}_{xx}\|&\|{s}_{xy}\|&0\\
    \|{s}_{yx}\|&\|{s}_{yy}\|&0\\
    -\|{p}_{zz}{\sigma}_{zx}\|&-\|{p}_{zz}{\sigma}_{zy}\|&-1\\
    \end{array}\right)^{-1} \no \\
    \times\left(\begin{array}{ccc}
    1&0&\|{s}_{xx}{\rho}_{xz}+{s}_{xy}{\rho}_{yz}\|\\
    0&1&\|{s}_{yx}{\rho}_{xz}+{s}_{yy}{\rho}_{yz}\|\\
    0&0&-\|{p}_{zz}\|\\
    \end{array}\right).
\end{eqnarray}
The inversion of the first matrix and the product of the two
matrices lead to the following expressions of the matrix elements
for the effective resistivity
\begin{eqnarray}\label{effres}
   \overline{\rho}_{xx}&=&\frac{1}{D}\|{s}_{yy}\|,\no\\
   \overline{\rho}_{xy}&=&-\frac{1}{D}\|{s}_{xy}\|,\no\\
   \overline{\rho}_{xz}&=&\frac{1}{D}\left(\|{s}_{yy}\|\|{s}_{xx}{\rho}_{xz}+{s}_{xy}{\rho}_{yz}\|\right.\no\\
   &&-\left.\|{s}_{xy}\|\|{s}_{yx}{\rho}_{xz}+{s}_{yy}{\rho}_{yz}\|\right),\no\\
   \overline{\rho}_{yx}&=&-\frac{1}{D}\|{s}_{yx}\|,\no\\
   \overline{\rho}_{yy}&=&\frac{1}{D}\|{s}_{xx}\|,\no\\
   \overline{\rho}_{yz}&=&\frac{1}{D}\left(\|{s}_{xx}\|\|{s}_{yx}{\rho}_{xz}+{s}_{yy}{\rho}_{yz}\|\right.\no\\
   &&-\left.\|{s}_{yx}\|\|{s}_{xx}{\rho}_{xz}+{s}_{xy}{\rho}_{yz}\|\right),\no\\
   \overline{\rho}_{zx}&=&\frac{1}{D}\left(\|{s}_{yx}\|\|{p}_{zz}{\sigma}_{zy}\|-\|{s}_{yy}\|\|{p}_{zz}{\sigma}_{zx}\|\right),\no\\
   \overline{\rho}_{zy}&=&\frac{1}{D}\left(\|{s}_{xy}\|\|{p}_{zz}{\sigma}_{zx}\|-\|{s}_{xx}\|\|{p}_{zz}{\sigma}_{zy}\|\right),\no\\
   \overline{\rho}_{zz}&=&\|{p}_{zz}\|+\frac{1}{D}\left((\|{s}_{yx}\|\|{p}_{zz}{\sigma}_{zy}\|-\|{s}_{yy}\|\|{p}_{zz}{\sigma}_{zx}\|)\right.\no\\
   &&\times\|{s}_{xx}{\rho}_{xz}+{s}_{xy}{\rho}_{yz}\|\,\no\\
   &&+(\|{s}_{xy}\|\|{p}_{zz}{\sigma}_{zx}\|-\|{s}_{xx}\|\|{p}_{zz}{\sigma}_{zy}\|)\no\\
   &&\times\left.\|{s}_{yx}{\rho}_{xz}+{s}_{yy}{\rho}_{yz}\|\right),
\end{eqnarray}
where we have introduced the denominator $ D \equiv
\|{s}_{xx}\|\|{s}_{yy}\| - \|{s}_{xy}\|\|{s}_{yx}\|$.
Eqs.~(\ref{effcond}) and (\ref{effres}) are valid in the general
case. We shall now give reduced expressions in some particular cases.

Without magnetization and for a cubic lattice, the symmetry
imposes $\sigma_{i\neq j}^{l_1l_2}=\rho_{i\neq j}^{l_1l_2}=0$. In
addition, we have $s_{ii}^{l_1l_2}=\sigma_{ii}^{l_1l_2}$ for
$i\in\{x,y\}$ and $p_{zz}^{l_1l_2}=\rho_{zz}^{l_1l_2}$ (see
Eq.~(\ref{inv})). Thus, the effective conductivity  and
resistivity tensors (Eqs.~(\ref{sys8}) and  (\ref{sys9}),
respectively) reduce to
\begin{eqnarray}\label{resm0}
    \overline{\sigma}&=&
    \left(\begin{array}{ccc}
    \|{\sigma}_{xx}\| & 0 & 0 \\
    0 & \|{\sigma}_{yy}\| & 0 \\
    0 & 0 & \frac{1}{{\|{\rho}}_{zz}\|}
    \end{array}\right),\\
    \label{resm0bis}\overline{\rho}&=&
    \left(\begin{array}{ccc}
    \frac{1}{\|{\sigma}_{xx}\|} & 0 & 0 \\
    0 & \frac{1}{\|{\sigma}_{yy}\|} & 0 \\
    0 & 0 & \|{\rho}_{zz}\|
    \end{array}\right).
\end{eqnarray} \\

When the magnetization is along the x-direction, we have
$\sigma_{ix}^{l_1l_2} = \sigma_{xi}^{l_1l_2} = \rho_{ix}^{l_1l_2}
= \rho_{xi}^{l_1l_2} = 0$ for $i\in\{y,z\}$ (and similarly for s
and p). They are no particular relations between
$s_{ij}^{l_1l_2}$ and $\sigma_{ij}^{l_1l_2}$ or between
$p_{ij}^{l_1l_2}$ and $\rho_{ij}^{l_1l_2}$. Thus, the effective
conductivity and resistivity tensors are equals to
\begin{eqnarray}\label{resmx}
    \overline{\sigma}&=&
    \left(\begin{array}{ccc}
    \|{s}_{xx}\| & 0 & 0 \\
    0 & \|{s}_{yy}\|-\frac{\|{s}_{yy}{\rho}_{yz}\|\|{p}_{zz}{\sigma}_{zy}\|}{\|{p}_{zz}\|} & -\frac{\|{s}_{yy}{\rho}_{yz}\|}{\|{p}_{zz}\|} \\
    0 & \frac{\|{p}_{zz}{\sigma}_{zy}\|}{\|{p}_{zz}\|} & \frac{1}{\|{p}_{zz}\|}
    \end{array}\right),
\end{eqnarray}
\begin{eqnarray}\label{condmx}
    \overline{\rho}&=&
    \left(\begin{array}{ccc}
    \frac{1}{\|{s}_{xx}\|} & 0 & 0 \\
    0 & \frac{1}{\|{s}_{yy}\|} & \frac{\|{s}_{yy}{\rho}_{yz}\|}{\|{s}_{yy}\|} \\
    0 & -\frac{\|{p}_{zz}{\sigma}_{zy}\|}{\|{s}_{yy}\|}
    &\|{p}_{zz}\|-\frac{\|{s}_{yy}{\rho}_{yz}\|\|{p}_{zz}{\sigma}_{zy}\|}{\|{s}_{yy}\|}
    \end{array}\right),
\end{eqnarray}
where $D = \|{s}_{xx}\|\|{s}_{yy}\|$.
The results with a magnetization along the y-direction can be obtained from Eqs.~(\ref{resmx}) and (\ref{condmx}) by exchanging x and y indices.

When the magnetization is along the z-direction, we have
$\sigma_{iz}^{l_1l_2} = \sigma_{zi}^{l_1l_2} = \rho_{iz}^{l_1l_2}
= \rho_{zi}^{l_1l_2} = 0$ for $i\in\{x,y\}$. In addition, we have
$s_{ij}^{l_1l_2}=\sigma_{ij}^{l_1l_2}$ for $\{i,j\}\in\{x,y\}$
and $p_{zz}^{l_1l_2}=\rho_{zz}^{l_1l_2}$ (see Eq.~(\ref{inv})).
Thus, the effective conductivity and resistivity tensors are
equals to
\begin{eqnarray}\label{resmz}
    \overline{\sigma}&=&
    \left(\begin{array}{ccc}
    \|{\sigma}_{xx}\| & \|{\sigma}_{xy}\| & 0 \\
    \|{\sigma}_{yx}\| & \|{\sigma}_{yy}\| & 0 \\
    0 & 0 & \frac{1}{\|{\rho}_{zz}\|}
    \end{array}\right),\\
    \overline{\rho}&=&
    \left(\begin{array}{ccc}
    \frac{1}{D}\|{\sigma}_{yy}\| & -\frac{1}{D}\|{\sigma}_{xy}\| & 0 \\
    -\frac{1}{D}\|{\sigma}_{yx}\| & \frac{1}{D}\|{\sigma}_{xx}\| & 0 \\
    0 & 0 & \|{\rho}_{zz}\|
    \end{array}\right),
\end{eqnarray}
where $ D = \|{\sigma}_{xx}\|\|{\sigma}_{yy}\| -
\|{\sigma}_{xy}\|\|{\sigma}_{yx}\|$. The fact that for CIP, the effective conductivity
$\overline{\sigma}_{ij}$ (where $\{i,j\}\in \{x,y\}$) is simply
equal to
$\|{\sigma}_{ij}\|=\sum_{l_1,l_2}{\sigma}_{ij}^{l_1,l_2}/L$ and
that for CPP, the effective
conductivity $\overline{\sigma}_{zz}$ is equal to
$1/\|{\rho}_{zz}\|=1/(\sum_{l_1,l_2}{\rho}_{zz}^{l_1,l_2}/L)$ has
been widely used in the literature\cite{Camblong,Zhang,Blaas}.
Assuming the local limit, it is thus possible to model the
multilayer as a network of resistors in series (in the CIP
geometry) or in parallels (in the CPP
geometry)\cite{Camblong,Valet}.\\

* Corresponding author; present address: Centre de Physique Th\'eorique, Luminy, Case 907, 13288 Marseille cedex 9, France; e-mail: crepieux@cpt.univ-mrs.fr

\end{multicols}
\end{document}